\documentclass[reprint,aps,prl,twocolumn,superscriptaddress,amsmath,amssymb,showpacs]{revtex4}
\usepackage{graphicx,esint,amsmath}
\usepackage{color}

\newcommand{\TW}{\ensuremath{\mathrm{TW/cm^2}}}

\begin{document}

\title{Femtosecond filamentation by intensity clamping at a Freeman resonance}

\author{Michael Hofmann}
\affiliation{Weierstra\ss-Institut f\"ur Angewandte Analysis und Stochastik, 10117 Berlin, Germany}

\author{Carsten Br\'ee}
\affiliation{Weierstra\ss-Institut f\"ur Angewandte Analysis und Stochastik, 10117 Berlin, Germany}
\affiliation{Max-Born-Institut f\"ur Nichtlineare Optik und Kurzzeitspektroskopie, 12489 Berlin, Germany}

\date{\today}

\begin{abstract}
We demonstrate that Freeman resonances have a strong impact on the nonlinear optical response in femtosecond filaments.
These resonances decrease the transient refractive index within a narrow intensity window and strongly affect the filamentation dynamics.
In particular, we demonstrate that the peak intensity of the filament can be clamped at these resonances, hinting at the existence of new regimes of filamentation with electron densities considerably lower than predicted by the standard model.
This sheds a new light on the phenomenon of filamentary intensity clamping and the plasmaless filaments predicted by the controversial higher-order Kerr model.
%
%
%
\end{abstract}

\maketitle

\noindent 
Femtosecond laser filaments are narrow beams of intense laser light in ionizing media which maintain their beamwaist over distances exceeding the linear diffraction length \cite{braun}.
The high optical intensities in filaments lead to transient refractive index modifications, and according to the established standard model of filamentation, they stem from an interplay of the all-optical Kerr effect and free electrons generated by multiphoton and tunneling ionization processes. 
Self-guided filaments are then understood to result from a balance of Kerr self-focusing and plasma defocusing \cite{standard_model}.
While the temporal evolution of a filamentary laser pulse is highly dynamic and subject to recurrent focusing and refocusing cycles \cite{replenishment}, a temporally averaged model can be shown to admit spatial soliton solutions, which produces the illusion of a nondiffracting beam \cite{aerosols,sulem}.

The standard model successfully fostered the theoretical understanding of various phenomena of nonlinear optics related to filamentation, like supercontinuum generation, pulse self-compression or terahertz generation \cite{supercontinuum,stibenz,tera}.
However, it  has recently been challenged by measurements of the cross-Kerr response in a pump-probe setup designed to detect the Kerr-induced birefringence \cite{loriot}. 
This revealed strong deviations from the linear intensity dependence of the Kerr response, and led to the proposal of an extended model including higher-order terms in the intensity. 
Above a threshold intensity, these terms turn the Kerr response into a defocusing nonlinearity, leading to the prediction of plasmaless filaments \cite{plasmaless}.
The higher-order Kerr effect (HOKE) model has been heavyly debated \cite{kolesik2010,Bree2011,Koehler2013,Bejot2013,Wahlstrand2011}.
Nevertheless, it turned out that some basic assumptions of the standard model should be reconsidered.
One of the major weaknesses of the standard model appears now to be the fact that it mixes perturbative and nonperturbative aspects of nonlinear optics in an inadmissible manner.
 While the nonlinear polarization density is treated perturbatively, the typical intensities in filaments exceed the validity range of perturbative multiphoton ionization models. 
Instead, in order to correctly describe ionization effects, some variant of the nonperturbative Keldysh theory \cite{Keldysh} is employed.
The inadequacy of a perturbative description of the optical response in filaments has been revealed in Ref.~\cite{Spott2014}.
Moreover, the standard model inherits the separate treatment of bound state and continuum response from the macroscopic Maxwell's equation.
However, it was shown that in the presence of a strong laser field, gauge variance renders the distinction between bound and continuum electrons ambiguous \cite{Koehler2013,Bejot2013}. 
Instead, bound states of free electrons, so called Kramers-Henneberger states, were proposed to contribute to HOKE \cite{Richter2013}.
Furthermore, the standard model assumes off-resonant excitation and neglects the dispersive character of the $\chi^{(3)}$ susceptibility. 
This issue was recently resolved in Ref.~\cite{Schuh2014}.

In the current Letter, we demonstrate that the optical response in filaments is nonperturbative and governed by transient atomic resonances, so called Freeman resonances \cite{Freeman1987}, whose impact on the optical response was previously indicated in Ref.~\cite{Richter2013}.
They occur when the AC Stark shift of atomic levels is of the order of a photon energy, which is the case for intensities exceeding some ten \TW. 
Analyzing their impact on filamentary propagation, we show that Freeman resonances support intensity clamping, giving rise to new regimes of filamentation.

In order to analyze the transient response properties of atomic hydrogen dressed by a strong pump pulse, we solve the time dependent Schr\"odinger equation (TDSE) for a commonly used 1D model atom \cite{Joachain2011},
\begin{equation}
  \label{eq:tdse}
  i\partial_t\psi=-\frac{1}{2}\partial_z^2\psi-\frac{1}{\sqrt{z^2+\alpha^2}}\psi+E(t)z\psi
\end{equation}
where $\alpha=\sqrt 2$ was chosen to match the ionization potential $I_p=13.6$\,eV ($0.5$au) of atomic hydrogen. 
The model atom is subject to a total electric field $E(t)=E_{\rm pu}(t)+E_{\rm pr}(t)$ with a strong pump pulse $E_{pu}$ and a weak probe $E_{pr}$ of identical carrier wavelengths $\lambda=800$\,nm. 
The optical response of the dressed atom is derived from the differential dipole response \cite{brown2012}
\begin{equation}
  \delta P[E_{pr}](t)=P[E_{pu}+E_{pr}](t)-P[E_{pu}](t),
\end{equation}
where
\begin{equation}
  P[E](t)=-\rho_{\rm nt}q_e\langle\psi(t)|z|\psi(t)\rangle
\end{equation}
is the polarization density, $\rho_{\rm nt}$ the atomic density and $q_e$ the electron charge.
We are interested in the cross-induced transient refractive index seen by the probe.
This is obtained by calculating the polarization response $\delta P(\omega_0)$ at the probe frequency, after removing the negative frequency contributions of the probe field.
%
This eliminates spurious contributions to $\delta P(\omega_0)$ due to multiwave mixing and is achieved by replacing the real probe field ${E}_{pr}$ with the complex analytic signal $\mathcal{E}_{pr} = E_{pr} + i \mathcal H( E_{pr} )$, where $\mathcal H$ is the Hilbert transform.
%
%
%
%
%
%
By solving the TDSE (\ref{eq:tdse}) seperately for the real and complex parts of $\mathcal{E}_{pr}$, the differential response $\delta P$ to the complex probe field can be evaluated according to
\begin{equation}
  \delta P[\mathcal{E}_{pr}](t)=\delta P[\mathrm{Re}\,\mathcal{E}_{pr}](t)+i\,\delta P[\mathrm{Im}\,\mathcal{E}_{pr}](t).
\end{equation}

The medium is dressed by a flat-top pump field of variable intensity $I$.
The field is switched on and off following a four-cycle $\mathrm{cos}^2$-envelope with a constant amplitude along 40 cycles inbetween. 
The weak complex probe field is 124 cycles flat-top pulse with a peak intensity of $I_{pr}=1\,\mathrm{W/cm^2}$. 
The transient susceptibility then reads as
\begin{equation}
  \chi(\tau)=\frac{\delta \tilde{P}(\omega_0,\tau)}{\epsilon_0\tilde{E}_{pr}(\omega_0,\tau)},
  \label{eq:chitrans}
\end{equation}
where $\omega_0$ denotes the carrier frequency of pump and probe. 
$\tilde{P}(\omega,\tau)$ and $\tilde{E}(\omega,\tau)$ are short-time Fourier transforms with a 20\,fs FWHM Gaussian window $w(t-\tau)$ centered at $t=\tau$.
The transient refractive index change is then obtained from 
\begin{equation}
  \Delta n(\tau)=\sqrt{1+\mathrm{Re}\,\chi(\tau)}-n_0,
\end{equation}
where $n_0$ is the field-free refractive index.
\begin{figure}[h!]
  \centerline{\includegraphics[width=\linewidth]{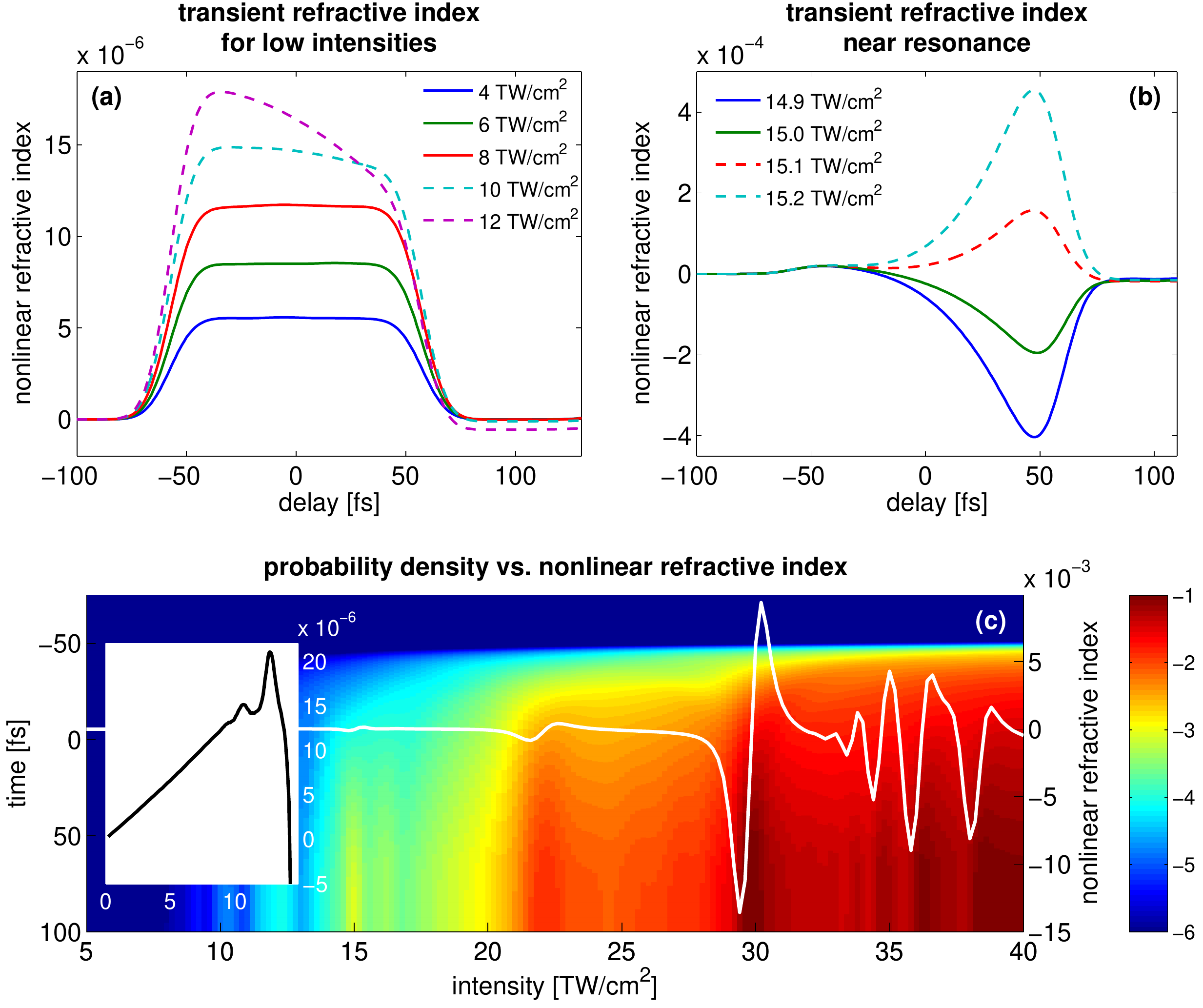}} 
  \caption{(color online) (a) Transient refractive index induced by a 40-cycle flat-top pump for low intensities and (b) in the vicinity of the first prominent Freeman resonance at $15$TW/cm$^2$. 
(c)  Logarithmic plot of probability density beyond 200 au. 
Solid line: nonlinear refractive index change. 
Inset: Closeup on low intensities.}
\label{fig:flattop}
\end{figure}
For weak pump intensities smaller than 10\,\TW, Fig.~\ref{fig:flattop}a) shows that $\Delta n(\tau)$  follows the intensity envelope of the pump pulse, in accordance with the Kerr model of the optical response based on an instantaneous $\chi^{(3)}$-nonlinearity.
As our grid is sufficiently large (6000\,au radius), we capture the full plasma contribution to the refractive index, which is evident from the constant negative $\Delta n(\tau)$ in the wake of the 12\,\TW\ pulse, cf.~Fig.~\ref{fig:flattop}a).
However, at this intensity, the decrease in $\Delta n(\tau)$ during the pulse is much larger than the plasma contribution in the wake of the pulse, a first evidence of deviations from the standard model.
In the vicinity of 15\,\TW\ as shown in Fig.~\ref{fig:flattop}b), these deviations are dramatic and indicate the breakdown of a perturbative, instantaneous description of the nonlinearity. 
In fact, in the trailing part of the pump in Fig.~\ref{fig:flattop}b), $\Delta n(\tau)$ exhibits a resonance pattern, with pronounced negative response (solid lines) below and positive response (dashed lines) above 15.05\,\TW.
In order to expose the origins of this behavior, we plot in Fig.~\ref{fig:flattop}c) the electron density leaving a spatial  range of 200\,au versus time and pump intensity.
This exposes a sharp peak in the electron density at 15\,\TW\ and further peaks at higher intensities.
These peaks are related to resonance enhanced multiphoton ionization (REMPI) due to the presence of Freeman resonances \cite{Freeman1987}.
They arise as the intensity is increased beyond a $K-$photon channel closure (CC) \cite{Potvliege} defined by $K\hbar\,\omega_0=I_p+U_p$, where $I_p$ is the field-free ionization potential of the atom and $U_p=E^2/4\omega^2$ (in atomic units) is the ponderomotive potential.
The white solid line in Fig.~\ref{fig:flattop}c) shows the pump-induced refractive index change, evaluated at the carrier frequency, i.e.\ $\Delta n(\omega_0)=\sqrt{1+\mathrm{Re}\,\chi(\omega_0)}-n_0$, where $\chi(\omega_0)$ is obtained from Eq.~(\ref{eq:chitrans}) using a constant window $w\equiv 1$.
%
%
While $\Delta n$ increases linearly for intensities below 10\,\TW, cf.\ the inset in Fig.~\ref{fig:flattop}c) , REMPI leads to sharp resonances for higher intensities.
%
Especially after the ten photon CC around 30\,\TW, $\Delta n$ varies by orders of magnitude, leading us to suspect that Freeman resonances dominate the optical response in this regime.

The sharpness of these resonances may also be attributed to the employed flat-top pulse, enabling the field-free eigenstates to adiabatically adapt to field-dressed eigenstates.
However, experimental femtosecond pulses usually exhibit a variable intensity envelope. 
We therefore repeat our calculations for a pulse with  $\cos^2$ shape, where the pulse duration of $\sim90$\,fs FWHM (96 cycles total) matches that of the original HOKE experiment \cite{loriot}. 
Figures 2a) and b) depict the transient refractive index in the low-intensity regime and in the vicinity of the resonance at 15\,\TW.
Qualitatively, we observe the same behavior as for the flat-top pump. 
However, due to the non-constant intensity envelope, the resonance appears smeared out, and we may expect the resonant features to disappear for even shorter pulses.   
%
%

\begin{figure}[h!]
\centerline{\includegraphics[width=\linewidth]{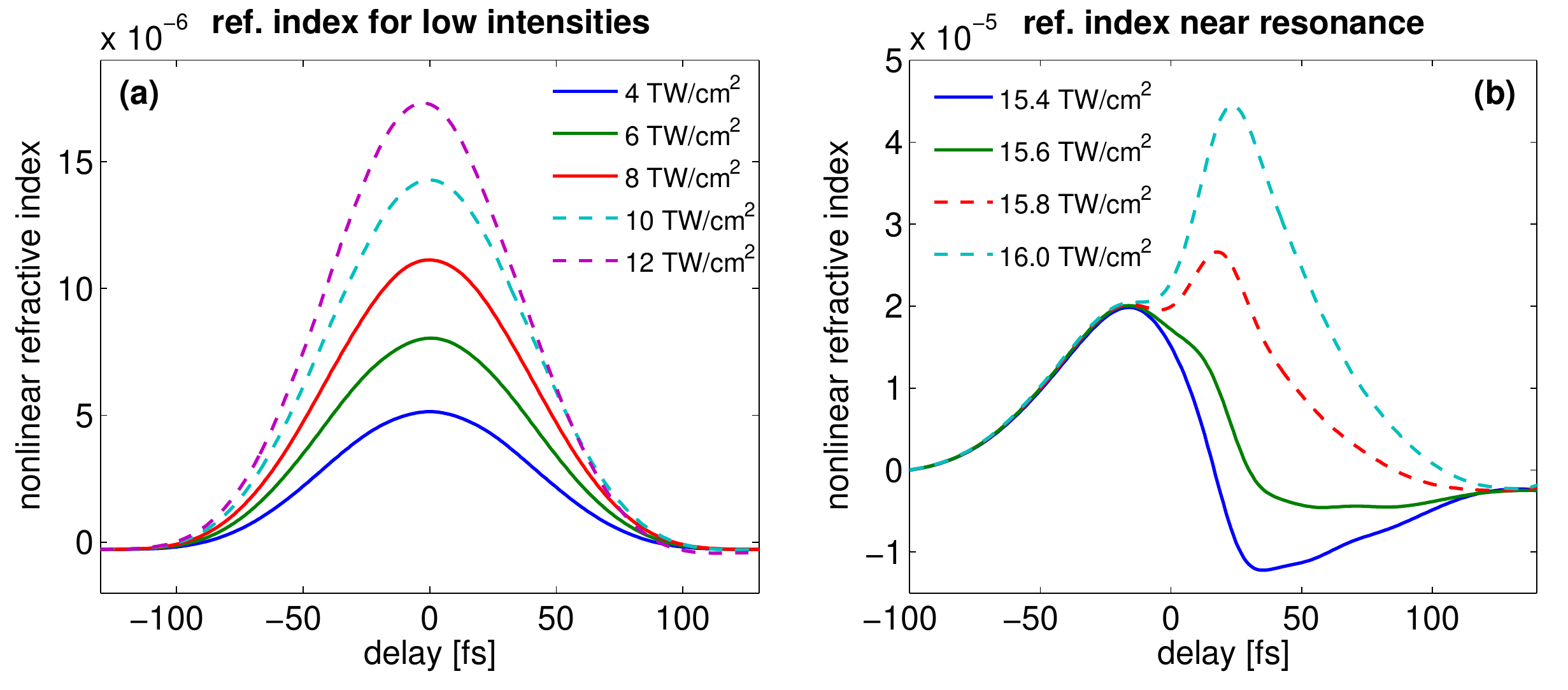}} 
\caption{ (color online) Transient refractive index as in Fig.~\ref{fig:flattop}a) and b), but for a 90fs $\cos^2$-pulse. 
The resonance and hence the refractive index saturation are still clearly visible.}
\label{fig:cos2}
\end{figure}

\begin{figure}[h!]
\centerline{\includegraphics[width=\linewidth]{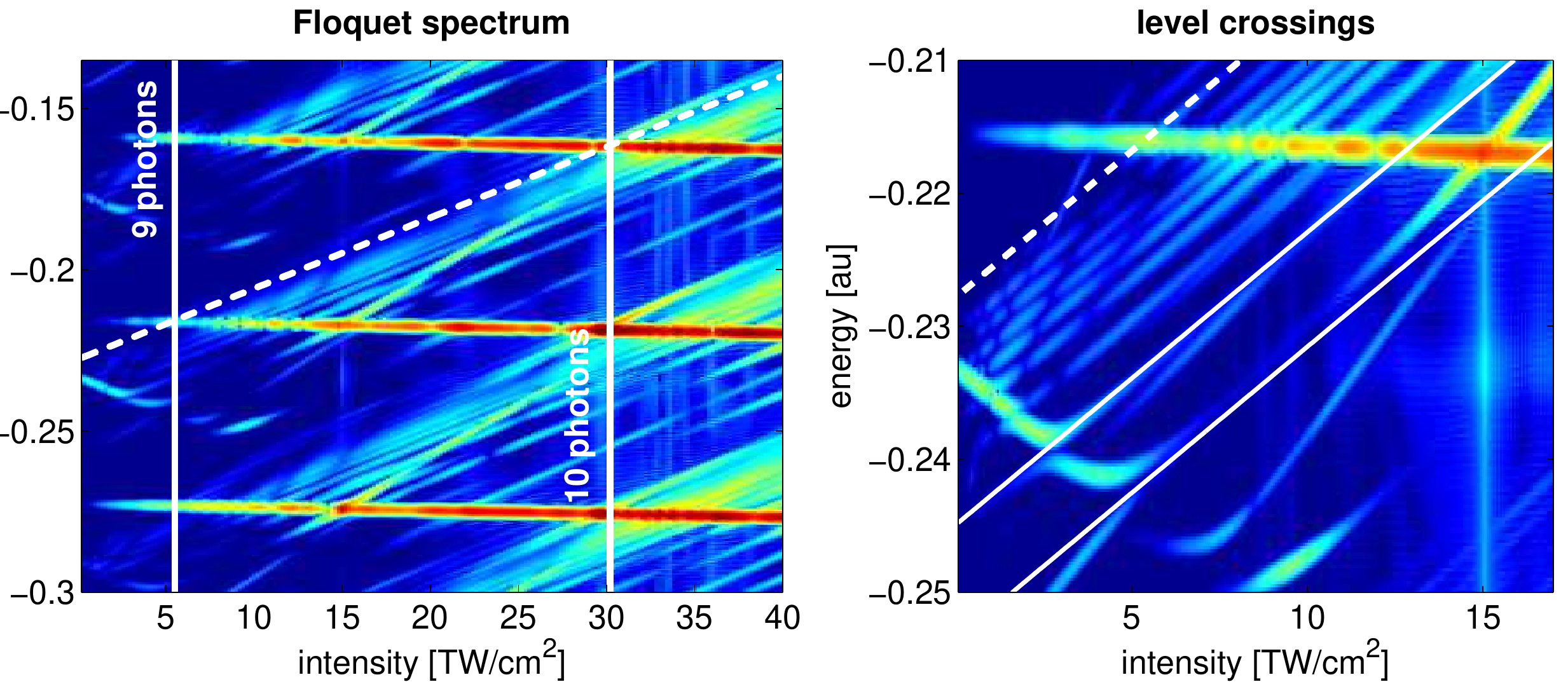}} 
\caption{ (color online) (a) Floquet spectrum versus intensity. 
Solid lines mark 9- and 10-photon CC, respectively.
(b) Closeup on the five-photon dressed ground state in the vicinity of the first excited state. 
Solid and dashed lines: ponderomotive upshift of 7th and 9th excited field-free states and continuum limit (bottom to top) in  $N=-4$ Floquet block.}
\label{fig:floquet}
\end{figure}

To further corroborate our hypothesis on the importance of Freeman resonances, we analyze the Floquet quasienergies using the method of Ref.~\cite{Floquet}.
Our results for pulses with 80 and 160 cycles flat-top are shown in Fig.~\ref{fig:floquet}a) and b), respectively.
The increased pulse durations do not qualitatively alter the result, but improve the sharpness and frequency resolution of the Floquet spectral peaks.
The prominent lines originating at $\omega=-0.5+N\omega_0$ correspond to the $N-$photon dressed ground state, where $\omega_0=0.057$, in atomic units. 
With increasing intensity, the ground state energy decreases only slightly. 
The dressed excited states are subject to the AC Stark shift, as visualized by the dashed white line in Fig.~\ref{fig:floquet}a) which shows the ponderomotively upshifted continuum limit in the $N=-4$ Floquet block.
Interestingly, Fig.~\ref{fig:floquet}a) exhibits level crossings of the dressed ground state with excited states.
These crossings occur just after the nine- and ten-photon CCs indicated by the solid vertical lines, and their position on the intensity axis matches that of the resonances observed in our numerical pump probe experiment, cf.\ Fig.~\ref{fig:flattop}c).

In Fig.~\ref{fig:floquet}b), a closeup of the level crossings up to 15\,\TW\ is shown.
Below the continuum threshold in the $N=-4$ Floquet block (dashed line), we recognize occupied Rydberg states shifting with $U_p$ which eventually cross the $N=5$ dressed ground state.
%
%
At 5\,\TW\ and below, avoided crossings of the first excited state in the $N=0$ block (originating at 0.233\,au) with the 7th and 9th excited state in the N=-4 block (solid lines) are evident, upon which the participating states interchange their role.
%
%
Therefore, the observed crossings below 15\,\TW\ facilitate nine-photon transitions, which confirms that the resonant behavior of the transient refractive index stems from Freeman resonances. 

\begin{figure}[h!]
  \includegraphics[width=\linewidth]{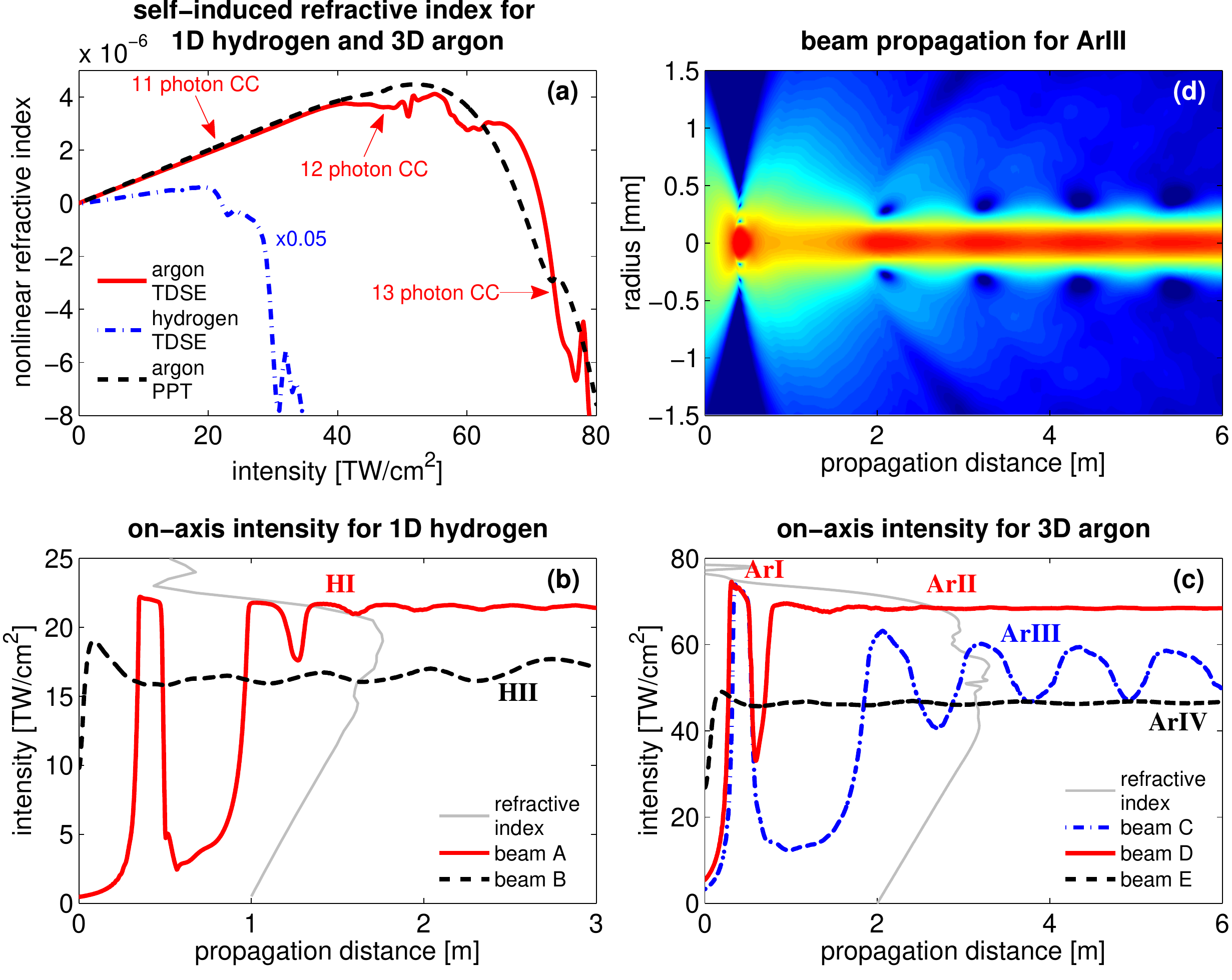}
  \caption{(color online) (a) Nonlinear refractive index $\Delta n(I)$ of hydrogen (dash-dotted line, reduced by factor 0.05) and argon (solid line). Dashed line: $\Delta n(I)$ according to standard model.
Arrows indicate channel closures in argon. 
(b) and (c) On-axis intensity versus propagation distance in hydrogen and argon, respectively, with initial beam parameters given in \ Table \ref{tab:beamdata}.
For comparison,  refractive index curves from panel a) are shown as thin grey lines.
(d) Evolution of the transverse beam profile for regime ArIII.}
  \label{fig:filament}
\end{figure}

These results raise the probing question whether the observed resonances have a notable influence on filament propagation.
To this purpose, we analyze the self-induced nonlinear refractive index (instead of the cross Kerr response of Eq.~(\ref{eq:chitrans})), which we obtain from 
\begin{equation}
  \chi(\omega_0)=\frac{\hat{P}[E_{pu}](\omega_0)}{\epsilon_0 \hat{E}_{pu}(\omega_0)}.
\end{equation}
For the 1D atomic hydrogen model, the resulting intensity dependent refractive index $\Delta n(I)$ is shown as the dash-dotted line in Fig.~\ref{fig:filament}a). 
In the low-intensity regime, it increases linearly with a slope $n_2=6.7\times 10^{-7}$cm$^2$/TW.
Due the onset of Freeman resonances, the refractive index exhibits deviations from the linear behavior for intensities beyond 15\,\TW.
However, the resonances appear less pronounced in the self-induced refractive index $n(I)$ than in its cross-induced counterpart $n^X(I)$.
This is a consequence of the relation $n^{X}(I)=n(I)+Idn/dI$ which generalizes the corresponding relation for the higher-order Kerr coefficients, $n^X_{2j}=(j+1)n_{2j}$ \cite{boyd}.

In order to increase the explanatory power of our approach, we extend our analysis to atomic argon by employing a 3D quantum model based on the single active electron approximation in an effective potential \cite{muller}.
The intensity dependent refractive index of argon is shown as the solid line
in Fig.~4a). 
We deduce a slope $n_2=1.06\times 10^{-7}$\,cm$^2/$TW in the low-intensity regime, in excellent agreement with the experimental value $n_2=0.98\times 10^{-7}$\,cm$^2/$TW \cite{lehmeier}.
Above the 12-photon CC at 47\,\TW, transient resonances locally decrease the refractive index.
The inversion intensity for which $\Delta n(I)$ changes its sign, amounts to $71.5$\,\TW, and a further transient resonance shows up slightly above the 13-photon CC at 73.6\,\TW.
Interestingly, the global behavior of the intensity dependent refractive index derived from the TDSE calculations for argon is in excellent agreement with the standard model which predicts a nonlinear index $\Delta n(I)=n_2 I-\rho/(2\rho_c)$. 
Here, we employ the measured value for $n_2$ according to \cite{lehmeier}.
The electron density generated by the $200$\,fs cos$^2$-pulse is denoted by $\rho$ and calculated according to the strong field ionization rate of Ref.~\cite{ppt}, and $\rho_c$ is the critical plasma density.
The resulting refractive index is shown as the dashed black line in Fig.~4a).
The inversion intensity according to the standard model is 68.3\,\TW, which is only slightly below the one derived from our TDSE calculations.
However, while the CC appear as local cusps in the standard model curve, we note the absence of transient resonances since the employed strong field ionization rate neglects excited bound states.

 A time averaged model equation for filamentary propagation then reads \cite{aerosols}
\begin{equation}
  \partial_z\mathcal{E}=\frac{i}{2k_0}\Delta_{\perp}\mathcal{E}+i\frac{\omega_0}{c}\Delta n(I)\mathcal{E}
  \label{eq:filament}
\end{equation}
where $\mathcal{E}$ is the complex electric field envelope, normalized as $I=|\mathcal{E}|^2$.
The carrier frequency is denoted as $\omega_0$, $n_0$ is the field-free refractive index, $k_0=n_0\omega_0/c$ is the wave-number at the carrier frequency, and $\Delta_{\perp}=\frac{1}{r}\partial_r r \partial_r$ is the transverse part of the Laplace operator, assuming cylindrical symmetry of the beam.
Depending on the chosen initial conditions, the beam explores different filamentation regimes, cf. Table \ref{tab:beamdata}, which compiles the beamwaist $w_0$, the initial peak intensity $I_0$, the ratio $P/P_{cr}$, where $P_{cr}=\lambda^2/2\pi n_0 n_2$ is the critical power for self-focusing, the focal length $f$ and the explored filamentation regime for the respective beam.
These regimes are distinguished by their respective clamping intensity and peak electron density.
In atomic hydrogen, we identify the plasma dominated filamentation regime HI, with a peak electron density of $4.9\times 10^{16}$\,cm$^{-3}$.
Remarkably, regime HII corresponds to subcritical intensity clamping at the 9-photon Freeman resonance in the vicinity of $15$\,\TW, cf.~the light grey line in Fig.~4b) depicting the nonlinear refractive index.
Here, the electron density amounts to $\rho=7.1\times 10^{15}$\,cm$^{-3}$, nearly an order of magnitude smaller than in HI.
In argon, we identify the plasma dominated regimes ArI and ArII, with electron densities of $1.9$ and $1.3\times 10^{17}$\,cm$^{-3}$, respectively.
Moreover, ArIII corresponds to intensity clamping at a Freeman resonance above the 12-photon CC, with a reduced electron density of $2.2\times 10^{16}$\,cm$^{-3}$.
For ArIII, the evolution of the radial beam profile versus propagation distance is shown in Fig.~4d). This reveals that the beam sheds radiation into its spatial surrounding upon converging to a spatial soliton solution \cite{sulem}.
In addition, we again observe subcritical intensity clamping (ArIV) with further reduced electron density $\rho=9.5\times 10^{15}$\,cm$^{-3}$.
\begin{table}[tbh]
\begin{tabular}{c|c|c|c|c|c}
Beam &  $w_0$(mm)    & $I_0(\TW)$ & $P/P{cr}$ & f(cm)& Regime   \\ \hline
A    & 1         & 0.46       & 3         & 50       & HI        \\
B    & 0.12      & 9.8        & 0.92      & $\infty$ & HII       \\
C    & 1         & 3.25       & 3         & 50       & ArI,ArII         \\
D    & 1         & 5.4        & 5         & 50       & ArI,ArIII        \\
E    & 0.2       & 26.8       & 1         & $\infty$ & ArIV       \\
\end{tabular}
\caption{Initial data for Gaussian beam propagation and filamentation regimes explored.}
\label{tab:beamdata}
\end{table}

In conclusion, we found that Freeman resonances strongly impact the optical response.
Since they are closely related to CCs, this sheds new light on the results of Ref.~\cite{Bejot2013} which made CCs accountable for the HOKE.
Our results imply the existence of new filamentation regimes due to intensity clamping at a Freeman resonance, especially for longer pulses of some $100$\,fs duration.
This is in accordance with a recent experimental result which demonstrated that for 200fs pulses undergoing filamentation a ``plasmaless'' postfilament regime \cite{mitrofanov} emerges.
However, while in former works, hypothetical plasmaless filaments were attributed to the HOKE, our research reveals a completely different underlying mechanism.
Moreover, our results shed an interesting perspective on subcritical intensity clamping \cite{novoa} and may have interesting implications with regard to bi- or multistable beam self-trapping \cite{kaplan}.
Finally, the presence of transient resonances implies that a HOKE-like phenomenological description of the refractive index in terms of an instantaneous power series in $I$ is inadequate, at least for the pulse durations considered here.
Regarding future research, note that the impact of Freeman resonances depends sensitively on the temporal intensity envelope of the pump, which exhibits a pronounced dynamics upon filamentary propagation \cite{replenishment}.
For an exact description, it is  therefore inevitable to replace our time-averaged model~(\ref{eq:filament}) with a filamentation model derived from first principles.

We thank Matthias Wolfrum, Oleh Omel'chenko, and G\"unter Steinmeyer for 
valuable discussions. 
Financial 
support by the Deutsche Forschungsgemeinschaft, grant BR 4654/1-1, is gratefully acknowledged.

\end{document}